\RequirePackage{rotating}
\documentclass[manuscript,screen]{acmart}

\def\BibTeX{{\rm B\kern-.05em{\sc i\kern-.025em b}\kern-.08emT\kern-.1667em\lower.7ex\hbox{E}\kern-.125emX}}

\begin{document}

\title{The Shift from Writing to Pruning Software: \\
A Bonsai-Inspired IDE for Reshaping AI Generated Code
}

\author{Raula Gaikovina Kula}
\affiliation{%
 \institution{Osaka University}
 \country{Japan}}
 \email{raula-k@ist.osaka-u.ac.jp}

\author{Christoph Treude}
\affiliation{%
  \institution{Singapore Management University}
  \country{Singapore}
}
\email{ctreude@smu.edu.sg}
 
\renewcommand{\shortauthors}{Kula, et al.}

%
\begin{abstract}

The rise of AI-driven coding assistants is a clear indication that building software will never be the same again. While AI coding assistants integrate into existing Integrated Development Environments (IDEs), their full potential remains underutilized.
One key challenge is that these AI assistants are prone to hallucinations, which can lead developers down decision paths that AI should not have the authority to make, sometimes even without the user’s consent. Furthermore, current static-file IDEs lack mechanisms to address critical issues such as tracking the provenance of AI-generated code and integrating version control in a way that aligns with the dynamic nature of AI-assisted development.
As a result, developers are left without the tools to systematically manage, refine, and validate AI-generated code, making it difficult to ensure correctness, maintainability, and trust in AI-assisted development. Existing IDEs treat AI-generated code as static text, offering limited support for managing its evolution, refinement, or multiple alternative paths.

Drawing inspiration from the ancient art of Japanese Bonsai gardening, which emphasizes balance, structure, and deliberate pruning, we propose a shift in how we think about IDEs, where AI is free to generate in its true nature -- unconstrained by traditional file structures -- allowing for a more fluid and interactive approach to code evolution. We introduce the concept of a Bonsai-inspired IDE, structured as a graph of generated code snippets and multiple code paths, giving developers control over reshaping AI-generated code to fit their needs.
We present the key principles of the Bonsai IDE and highlight challenges, along with seven research directions aimed at making the Bonsai IDE a reality.
Our vision lies in the fundamental shift that involves moving away from a static file-based model toward an interactive, evolving system of generated, refined, and alternative code paths, where the IDE becomes a dynamic programming environment that co-evolves with AI-powered modifications rather than simply serving as a place to write and edit code.

\end{abstract}

\maketitle

\section{Introduction}

\begin{quote}
    \textit{“The art of bonsai is a journey of discovery, a path that leads to a profound understanding of the delicate balance between nature and human creativity.\footnote{\url{https://bonsaibotanica.com/the-science-behind-bonsai-natures-living-art/}}”}
\end{quote}

Modern Integrated Development Environments (IDEs) were built for human-written code, where developers control edits, revisions are manually structured, and changes follow predictable workflows. However, AI-generated code emerges dynamically, lacks inherent traceability, and often requires iterative refinement -- challenges that existing IDEs are not equipped to handle. Unlike human-authored code, AI-generated code can emerge unpredictably, lacks inherent provenance tracking, and integrates poorly with structured development workflows. As a result, developers struggle with AI hallucinations, where models generate plausible but incorrect code that is difficult to verify. Additionally, without clear version tracking, integrating and refining AI-generated code requires extensive manual effort, increasing the risk of introducing subtle errors. As the era of Generative AI progresses, there are growing discussions about shifts in the future of software development. One prominent perspective is that the focus will shift from the code itself to the intent behind the code's generation. 

Existing IDEs treat AI-generated code as static text, providing little support for managing its evolution. Current implementations, such as GitHub's Copilot\footnote{\url{https://github.com/features/copilot}}, OpenAI's ChatGPT\footnote{\url{https://openai.com/chatgpt/overview/}}, Google's Gemini\footnote{\url{https://gemini.google.com/}} and so on, position AI as an assistant or as an agent that works side-by-side with the developer. However, we argue that these approaches are vulnerable to hallucinations \cite{hallucinationsZiwei}, where AI generates code that can mislead the developer. Such hallucinations often involve reasoning, chain-of-thought, or deep search processes \cite{DBLP:cotreason} that are driven by AI but remain opaque to the developer.
A 2025 report indicated that closed AI models do not disclose their technical underpinnings\footnote{\url{https://responsibleinnovation.org/wp-content/uploads/2025/01/ARI-Report-Transparency-in-Frontier-AI.pdf}}.
Such examples of little transparency and traceability can lead developers down unintended paths of losing agency, making it challenging to trust AI-generated code without fully understanding the reasoning behind the decisions and whether it aligns with developer intentions. This lack of transparency forces developers to rely on AI suggestions without fully understanding the reasoning behind them, increasing the risk of integrating erroneous or suboptimal code. Without fine-grained control, developers may find it challenging to guide AI-generated code toward maintainable, high-quality solutions.

Current approaches attempt to retrofit AI into traditional IDEs, but these tools remain fundamentally designed around human-written code. Instead of forcing AI into outdated workflows, we propose a new paradigm: a Bonsai-inspired IDE that treats AI-generated code as an evolving structure, allowing developers to guide and refine it dynamically. Developers need an environment where AI-generated code is not merely inserted but actively managed, explored, and refined. In this vision paper, we introduce a novel concept that reimagines how generative AI technologies can be fully harnessed while ensuring that developers retain full agency over the generated code. The central idea is to shift away from viewing code as static, file-based structures and instead treating it as a dynamic collection of code snippets that are linked to the prompts from which they were generated. Drawing inspiration from Code Bubbles \cite{codebubbles}, which also moves away from file-based structures to code fragments, we propose a framework where, like a Bonsai tree that is meticulously pruned to achieve balance and aesthetic appeal, the code evolves and refines itself through careful, intentional modifications. This approach emphasizes flexibility and adaptability, allowing developers to shape and prune the codebase in alignment with their evolving goals and AI guidance, without losing control over the decision-making process.

The Bonsai-inspired IDE integrates features that embody the principles of Bonsai cultivation, emphasizing continuous evolution, adaptability, and thoughtful refinement throughout the development process.
IDE features can be aligned with bonsai styling principles to enhance the development experience. Asymmetry is reflected in the Code Generation Graph and Prompt-Driven Development View, where code evolves in a dynamic, non-linear fashion, mimicking organic growth. Simplicity is achieved through the Intent-Based Project Structure and AI-Assisted Code Morphing, focusing on essential interactions and minimizing complexity. Proportion is emphasized in the way Dynamic Code States and Hierarchical Generation Layers are presented, offering a balanced view of code at different stages of development. Lastly, Depth is integrated through the Interactive Code Evolution Timeline, Regeneration Networks, and AI Code Sandboxing, providing a rich and layered understanding of the code’s evolution, allowing developers to track, assess, and refine their work over time.

This vision redefines software development as a process of guided evolution rather than static editing. Instead of modifying static files, developers engage with AI-generated code as an evolving system -- branching, refining, and merging alternatives dynamically in response to changing requirements. The IDE becomes a collaborative space where human intent and AI generation co-evolve, shaping the codebase through continuous iteration.
The IDE transitions into a platform that co-evolves with AI modifications, allowing for a flexible, interactive approach to development. The research agenda outlined here focuses on key challenges such as maintaining provenance and tracking code evolution, replacing traditional file structures with prompt-driven navigation, ensuring fine-grained control over AI-generated code, and managing multiple generations and parallel explorations. Other challenges include integrating AI into continuous regeneration pipelines, ensuring the safety and performance of AI-generated code, and addressing cognitive load while collaborating with AI. The goal is to create a system that allows for seamless interaction, continuous refinement, and thoughtful decision-making, ensuring that AI-assisted coding evolves along with developer intentions. 

\section{Disruption of the IDE with AI Assistants}

In this section, we examine traditional IDEs and how early AI integrations introduced the concept of AI as a pair programmer or copilot.

\subsection{The Traditional IDE (Eclipse and VSCode)}

Traditional Integrated Development Environments (IDEs) such as Eclipse\footnote{\url{https://eclipseide.org/}} and Visual Studio Code (VSCode)\footnote{\url{https://code.visualstudio.com/}} have revolutionized the way developers approach coding. They provide developers with powerful tools that facilitate modular coding practices, from writing code at various levels (classes, methods, variables) to refactoring, building, and testing. Their user-friendly interfaces further enhance developer productivity by offering customization and intuitive navigation, making them indispensable tools in the modern software development landscape.
Eclipse, known for its extensive plugin support, offers a modular approach to development, allowing developers to integrate various tools such as debuggers, version control systems, and automated testing frameworks. VSCode, built on top of the popular Node.js platform, also caters to this modular coding paradigm by enabling the use of extensions that enhance functionality.

Developers leverage these IDEs for their ability to handle complex projects through features like code editing with syntax highlighting and auto-completion. They can refactor code by restructuring modules or classes, ensuring a clean and maintainable codebase. The build process is streamlined, often automated through tools integrated into the IDE, that compile and link different parts of the code together seamlessly. Additionally, testing frameworks within these environments allow developers to run test cases efficiently, ensuring robust functionality.
The user interface (UI) of traditional IDEs like Eclipse and VSCode is designed with developer productivity in mind. These interfaces typically consist of a sidebar for file navigation, a top menu bar for accessing commands, and an editor area where code is written and edited. Customization options allow developers to tailor the UI to their preferences, improving ease of use.
VSCode's interface, for instance, is known for its flexibility, with a market of thousands of extensions available to add functionality. Eclipse also offers significant customization, though it may be less user-friendly than VSCode in some aspects. Both IDEs prioritize a clean and intuitive layout that simplifies navigation and access to tools.
The modular coding approach supported by these IDEs is further enhanced by their UI design, which organizes code into logical sections, making it easier for developers to manage and manipulate different parts of their projects. This structure not only promotes efficient coding, but also aids in debugging and refining the codebase as needed\footnote{\url{https://code.visualstudio.com/docs}}.

\subsection{AI in Traditional IDEs: Pair Programming in a Static World}
In recent years, there has been a significant shift in software development practices with the integration of AI-driven assistants like GitHub Copilot into traditional IDEs such as VSCode. These tools are revolutionizing how developers approach their work by offering intelligent support at every stage of the coding process. One of the most notable features is its inline code suggestions, which provide real-time assistance to developers as they write or iterate on their code. This feature not only completes lines of code but also offers contextually relevant suggestions based on the project's requirements and best practices.
The chat interface within Copilot enables developers to engage in direct conversations with AI to generate or refactor source code. For example, a developer can prompt Copilot to write a function or refactor a loop into a more efficient structure, and AI will respond accordingly. This approach is especially useful for generating documentation comments or unit tests, producing contextually relevant outputs that improve project clarity.
Beyond simple auto-completion, Copilot offers real-time suggestions based on the developer's current context within VSCode. During coding, it automatically highlights or suggests code snippets tailored to the project structure and coding style, ensuring consistency across large-scale projects.
A standout feature of Copilot is its ability to act as an AI pair programmer, offering autocomplete-style suggestions in a conversational format. Developers can begin writing a line of code, and AI will propose completions that simulate a dialogue between two programmers, fostering collaborative problem solving.

According to GitHub\footnote{\url{https://github.blog/news-insights/research/research-quantifying-github-copilots-impact-on-code-quality/}}, research indicates that 85\% of developers felt more confident in their code quality when using GitHub Copilot and GitHub Copilot Chat. Additionally, code reviews became more actionable and were completed 15\% faster. Finally, 88\% of the developers reported that GitHub Copilot Chat helped them maintain a flow state by keeping them focused, reducing frustration, and increasing their overall enjoyment of coding.

\section{Vision: A Bonsai-Inspired IDE for Reshaping AI Generated Code}

\begin{quote}
   \textit{ “The art of bonsai is a living sculpture, shaped by the hands of the artist and the forces of nature.” – Unknown}
\end{quote}

The vision of an AI-native IDE is inspired by the cultivation of a Bonsai tree -- transforming software development from a static, file-based process into a dynamic, evolving system where AI generation and developer-guided refinement work in tandem.

\subsection{Intent-driven Navigation Beyond Files}
Like a Bonsai tree that grows under careful and intentional guidance, this AI-native IDE enables developers to navigate their codebase dynamically -- moving beyond rigid file structures to an intent-driven system based on semantic queries.
Developers can explore their code dynamically, using intent-based prompts and semantic queries to retrieve, modify, and refine generated code. This metaphor of a Bonsai tree emphasizes flexibility and growth, as developers prune and shape their codebase through interactions that are context-driven, rather than confined to the traditional file and directory structure.

\subsection{Parallel Code path Branching for AI Exploration}
Parallel AI exploration and branching in the AI-native IDE resemble the growth of a Bonsai tree, where multiple code paths evolve simultaneously and can be selectively refined. Rather than following a single, linear iteration, developers can branch out and explore alternative implementations in parallel, much like a Bonsai artist deciding how each branch will grow and evolve. The system maintains contextual awareness across branches, allowing developers to selectively merge, refine, or prune AI-generated code paths. This process reflects the Bonsai artist’s ability to shape multiple branches simultaneously, maintaining harmony and balance in the overall structure of the tree while adapting it to new growth opportunities. 

\subsection{Incorporating Bonsai Aesthetic Principles}
The art of bonsai styling employs various techniques to mould the tree’s growth and shape. Pruning, wiring, and pinching are used to achieve balance, proportion, and harmony. The aim is to replicate the presence of a mature tree in nature, imbuing it with character and presence.
The IDE design is guided by four key Bonsai styling principles, drawing on established sources on its aesthetics and methodology \cite{chan2014bonsai,chan2018bonsai, tomlinson1995complete}. 

\begin{enumerate}
    \item \textbf{Asymmetry:} Asymmetry is a core principle in Bonsai styling, where trees are shaped to reflect natural, organic growth rather than rigid symmetry. 
    \begin{enumerate}
        \item \textit{Code Generation Graph:} The IDE can display a graph that does not follow a rigid hierarchical structure but rather mimics the organic growth of a tree, where nodes branch out in a natural, non-symmetrical way. Linkages between generated code snippets and the prompts are dynamically visualized, rather than following strict parent-child hierarchies like traditional file trees. This approach highlights the flexibility and continuously evolving nature of AI-generated code.
        \item \textit{Prompt-Driven Development View:} Similar to asymmetry, prompts and modifications could be arranged in a more organic flow. The timeline or canvas would not necessarily be linear but would evolve dynamically as developers add new inputs or make changes, reflecting a more natural progression that is not bound by strict order.
    \end{enumerate}
    
    \item \textbf{Simplicity:} Simplicity is another crucial principle, as the beauty of bonsai lies in its simplicity, with each element carefully chosen to contribute to the overall composition. 
    \begin{enumerate}
        \item \textit{Intent-Based Project Structure:} Rather than overloading users with unnecessary information or complexity, the project structure focuses on developer intent, much like how bonsai pruning strips away unnecessary elements to showcase simplicity. The interface should be intuitive and minimalistic, focusing on the purpose of each AI interaction or constraint while hiding the complexity of the underlying code generation process.
        \item \textit{AI-Assisted Code Morphing:} Instead of a cluttered set of tools or settings, the code morphing process should be simple, focusing solely on the key constraints that affect code readability, efficiency, or style. The interface should allow users to seamlessly adjust these elements, maintaining a clean design that emphasizes ease of use.
    \end{enumerate}
    
    \item \textbf{Proportion:} Proportion plays a key role, ensuring that the size of the tree, the pot and any accompanying elements are in harmony, creating a sense of balance and visual appeal. 
    \begin{enumerate}
        \item \textit{Dynamic Code States:} The concept of dynamic code states needs to have proportionate interactions between versions of code. Switching between AI-generated alternatives should feel natural, with developers able to focus on relevant states without feeling overwhelmed by unnecessary versions. Each state -- raw AI generation, refinement, and finalized code -- should align with the project's development stage, providing sufficient context without overwhelming the user.
        \item \textit{Hierarchical Generation Layers:} The layers of code (base, refinement, and final version) should be represented proportionally in the IDE, with clear delimitations between each stage. Users should be able to easily toggle between these layers to get the right view of their work at any given time, without one layer feeling more dominant than the others.
    \end{enumerate}
    
    \item \textbf{Depth:} Finally, depth is incorporated by arranging branches and foliage in layers, giving the illusion of a larger tree and enhancing the overall aesthetic. 
    \begin{enumerate}
        \item \textit{Interactive Code Evolution Timeline:} The timeline should provide depth by allowing users to view a history of their project’s evolution. It could show layers of changes over time, allowing users to backtrack and see the progress from the original AI-generated code to the most refined version. This timeline could be visualized as a layered path, offering depth and context, as developers can see how each iteration influenced the code.
        \item \textit{Regeneration Networks:} These networks can add depth by visualizing how different prompts and modifications lead to divergent paths, giving developers a clear view of how one prompt or change can alter the entire project. This view would allow them to explore the depth of possibilities and refine their code with an understanding of the larger picture.
    \end{enumerate}
\end{enumerate}

\section{Research Challenges and Directions for an AI-Native IDE}
As AI-driven coding evolves, fundamental research questions arise about its integration, usability, and long-term impact on software engineering practices. This research agenda outlines seven critical challenges in AI-driven software development, from code provenance tracking to human-AI collaboration. In the following sections, we will explore the research directions, identify the associated challenges, and outline potential research topics that could advance the field. These research directions aim to tackle key challenges in AI-assisted software development, focusing on developer workflows, code quality, and human-AI collaboration.

The first research area focuses on the provenance and evolution tracking of AI-generated code. As AI technologies are increasingly used to assist in code generation, it becomes essential to track the origin and evolution of each code snippet. This involves maintaining a clear history of how and why specific pieces of code were generated, including the evolution of the AI model itself. One of the key challenges here is to determine how to maintain traceability without overwhelming developers with complex tracking systems.

\begin{enumerate}

    \item \textbf{Provenance and Code Evolution Tracking:} 
    Tracking the origin and evolution of AI-generated code is essential to maintain traceability and a clear development history. This research addresses the challenges of tracking provenance and integrating AI-driven version control systems.
    \begin{itemize}
        \item \textbf{Challenges:} 
        \begin{itemize}
            \item Maintaining precise provenance of how and why each snippet of code was generated, while efficiently storing and retrieving prompt-based code evolution graphs without overwhelming developers.
            \item Managing traceability between prompt modifications, AI refinements, and code changes.
        \end{itemize}
        \item \textbf{Research Topics:} 
        \begin{itemize}
            \item AI-driven version control: How should AI-generated code integrate into Git workflows or alternatives, and how can prompt lineage tracking algorithms be developed to track and visualize prompt-to-code transformations?
        \end{itemize}
    \end{itemize}

    \item \textbf{Intent-Based Code Navigation and Organization:} 
    Traditional file-based structures impose constraints on AI-generated code by enforcing rigid hierarchies. This research explores how to replace them with more flexible, intent-driven navigation systems.
    \begin{itemize}
        \item \textbf{Challenges:} 
        \begin{itemize}
            \item Replacing the file-system-based structure with a prompt-driven, context-aware navigation system while ensuring developer intuition aligns with non-linear, AI-generated structures.
            \item Creating a semantic representation of codebases that allows flexible retrieval and reorganization.
        \end{itemize}
        \item \textbf{Research Topics:} 
        \begin{itemize}
            \item Investigating how code can be structured as an evolving graph instead of static files, and how developers can efficiently query, browse, and refine AI-generated code.
            \item AI-assisted refactoring: Automatically restructuring projects based on evolving intent.
        \end{itemize}
    \end{itemize}

    \item \textbf{Interactive Regeneration and Constraints:} 
    To ensure that AI-generated code meets the developer's intent, the regeneration process must support fine-grained constraints and iterative refinements.
    \begin{itemize}
        \item \textbf{Challenges:} 
        \begin{itemize}
            \item Allowing fine-grained constraints on AI-generated code while still preserving developer control, while ensuring consistency when multiple AI-generated edits modify overlapping sections of the codebase.
            \item Creating an interactive, intuitive UI for developers to guide AI refinement dynamically.
        \end{itemize}
        \item \textbf{Research Topics:} 
        \begin{itemize}
            \item Constraint-based AI generation: How can we introduce semantic constraints into AI-assisted coding, allowing developers to modify AI-generated code at different abstraction levels?
            \item Context-aware regeneration: Developing AI models that adapt edits to fit existing project styles.
        \end{itemize}
    \end{itemize}

    \item \textbf{Multiple Generations and Parallel Exploration:} 
    AI can generate multiple solutions to the same problem, but selecting, comparing, and integrating these alternatives poses a significant challenge.
    \begin{itemize}
        \item \textbf{Challenges:} 
        \begin{itemize}
            \item Providing mechanisms to explore alternative AI-generated implementations while preventing decision fatigue for developers and avoiding fragmentation when multiple AI-generated versions are tested.
            \item Providing context-aware comparisons between multiple generations.
        \end{itemize}
        \item \textbf{Research Topics:} 
        \begin{itemize}
            \item Multi-objective code generation: How can we generate diverse solutions optimized for different objectives (performance, readability, security)?
            \item AI-assisted decision-making: Developing tools to rank, evaluate, and merge alternative implementations while supporting parallel AI branching in a version control system.
        \end{itemize}
    \end{itemize}

    \item \textbf{Continuous Regeneration Pipelines:} 
    The evolving nature of software demands that AI-generated code remain up-to-date as new dependencies and technologies emerge.
    \begin{itemize}
        \item \textbf{Challenges:} 
        \begin{itemize}
            \item Developing mechanisms to update AI-generated code dynamically in response to evolving dependencies, libraries, and best practices, while balancing automation with developer oversight to avoid excessive or unwanted code changes.
            \item Avoiding unintended regressions when AI re-generates parts of the codebase.
        \end{itemize}
        \item \textbf{Research Topics:} 
        \begin{itemize}
            \item AI-powered continuous integration: How should AI-generated code integrate into CI/CD pipelines?
            \item Evolutionary AI: Developing AI systems that self-update, maintaining software over time while enabling human-in-the-loop AI regeneration.
        \end{itemize}
    \end{itemize}

    \item \textbf{AI Code Sandboxing and Safety:} 
    Verifying AI-generated code for security vulnerabilities, performance issues, and compliance risks before integration into production systems.
    \begin{itemize}
        \item \textbf{Challenges:} 
        \begin{itemize}
            \item Ensuring AI-generated code is safe, secure, and performs well before integrating it, and automatically detecting security vulnerabilities, unsafe coding patterns, and performance bottlenecks.
        \end{itemize}
        \item \textbf{Research Topics:} 
        \begin{itemize}
            \item Automated AI code verification: How can we sandbox and test AI-generated code before execution?
            \item Legal and ethical AI: How do we ensure AI-generated code adheres to licensing and ethical guidelines?
        \end{itemize}
    \end{itemize}

    \item \textbf{Human-AI Collaboration and Cognitive Load:} 
    As AI tools integrate more into development workflows, reducing cognitive load is essential to avoid overwhelming developers with excessive choices.
    \begin{itemize}
        \item \textbf{Challenges:} 
        \begin{itemize}
            \item Avoiding cognitive overload when presenting too many AI-generated options while designing interfaces that balance automation with developer control.
            \item Developing mechanisms to improve developer trust and understanding of AI-generated code through transparency and explainability.
        \end{itemize}
        \item \textbf{Research Topics:} 
        \begin{itemize}
            \item Human factors in AI: How does AI-assisted coding affect cognitive load and decision-making?
            \item Explainable AI in SE: How can we make AI-generated code interpretable and trustworthy while adapting AI assistance to individual developers' coding styles and expertise levels?
        \end{itemize}
    \end{itemize}

\end{enumerate}

\section{Summary}

\begin{quote}
    \textit{“The art of the Bonsai IDE is a journey of discovery, a path that leads to a profound understanding of the delicate balance between AI and human creativity in software engineering.”}
\end{quote}

\bibliographystyle{ACM-Reference-Format}
\bibliography{2030-bonsai}

\end{document}